\documentstyle[floats,aps]{revtex}

\begin{document}
\draft

\twocolumn[\hsize\textwidth\columnwidth\hsize\csname
@twocolumnfalse\endcsname

\title {\bf Transferable relativistic Dirac-Slater pseudopotentials}
\author{Ilya Grinberg, Nicholas J. Ramer and Andrew M. Rappe}
\address{Department of Chemistry and Laboratory for Research on the
Structure of Matter\\ University of Pennsylvania, Philadelphia, PA
19104} 
\date{\today} 
\maketitle

\begin{abstract} 

We present a method for constructing a scalar-relativistic pseudopotential which provides  exact agreement with relativistic Dirac-Slater all-electron eigenvalues at the reference configuration.  All-electron wave functions are self-consistently computed in the valence region at the exact all-electron scalar relativistic eigenvalues.  This method improves transferability of the resulting pseudopotential and presents a better starting point for the designed non-local pseudopotential approach {[Phys. Rev. B {\bf 59}, 12471 (1999)]}.  We present calculations for the gold atom as an example of the new  approach.

\end{abstract} 
]

$Ab\; initio$  density functional theory~\cite{HK,KS} (DFT) calculations have been used extensively over the past thirty years to study metals, insulators and semiconductors.  The plane-wave pseudopotential method has been widely  used because it is  both accurate and fast, enabling calculations on large systems inaccessible with other methods.  The pseudopotentials in these calculations mimic the effect of the nuclei  and core electrons  on the valence electrons, dramatically reducing the computational cost of the calculations.  However, poorly constructed pseudopotentials will lead to inaccuracies in the solid-state results.  The most important indicator of the quality of the pseudopotential is transferability, the ability of the pseudopotential to reproduce the all-electron valence eigenvalues of various atomic configurations.  In addition, pseudopotentials requiring a smaller basis set can significantly speed up the solid-state calculation.  

Several methods have been developed to accomplish these goals.  The $ab$ $initio$ pseudopotentials of Hamann, Schl\"uter and Chiang~\cite{HSC}  employ a different spherically-symmetric potential for each angular momentum, which  allows   enforcement of the norm-conservation condition at one atomic configuration, the reference configuration.  Norm conservation guarantees exact agreement between the all-electron and pseudopotential eigenvalues and wave functions beyond the core radius $r_c$ for the reference configuration,  and it greatly improves transferability in other configurations.  The semilocal norm-conserving pseudopotentials were  further developed by Kleinman and Bylander~\cite{KB} who transformed the pseudopotential into fully separable nonlocal form, dramatically reducing the memory cost.  The optimized pseudopotential method developed by Rappe $et$ $al.$~\cite{RappePS} reduces the plane-wave cutoff necessary to use norm-conserving pseudopotentials. The designed nonlocal (DNL) approach of Ramer and Rappe~\cite{RRPS} significantly improves the transferability without affecting the convergence properties.  Goedecker $et$ $al.$~\cite{GTH} have developed  highly transferable dual-space multiple-projector pseudopotentials  which are  expressed as sums of Gaussians.  The Vanderbilt ultrasoft pseudopotential method~\cite{Vdblt}  further lowers the plane-wave cutoff by discarding norm conservation but restoring the correct charge with augmentation charge density functions.  In this approach, good transferability is achieved by the use of several projectors.  The combination of ultrasoft pseudopotentials with  Bl\"ochl's projector-augmented wave~\cite{Blochl} method can provide extremely good convergence and transferability.    A feature common to all these methods is the preservation of  exact agreement of the all-electron and valence eigenvalues  at the reference configuration.
	
Many technologically important materials contain elements with $Z>54$.  For such heavy atoms, relativistic effects cannot be ignored without severe consequences for the accuracy of the calculations. The most important  change to the Kohn-Sham Hamiltonian is in the kinetic-energy operator. In DFT all-electron atomic calculations the  Dirac-Slater and the Koelling-Harmon~\cite{KH} approaches are widely used. The Dirac-Slater equation includes spin-orbit coupling,  producing (for $s=\frac{1}{2}$) a pair of spin-dependent (up and down) wave functions and eigenvalues for every orbital with $l>0$.  The Koelling-Harmon approach omits the spin-orbit interaction from the Hamiltonian but retains all other relativistic kinematic effects. Therefore a  single  ``spin-averaged'' wave function and eigenvalue is produced for each atomic orbital.  This is less accurate than the Dirac-Slater method, but the error thus introduced is often acceptably small.   In addition to kinetic and spin-orbit effects, the exact relativistic  correction for the local density approximation (LDA) exchange functional  is also known but has a  small effect on the valence states.   Recently the relativistic exchange functional has been derived within the  generalized gradient approximation~\cite{EKD}.

The direct relativistic effects on the valence states of a  heavy atom are small.  Relativity strongly affects the core states, changing the self-consistent potential seen by the valence orbitals.  This implies that a pseudopotential can effectively incorporate relativistic effects in non-relativistic solid-state pseudopotential calculations.  With a Koelling-Harmon reference atomic all-electron calculation,  the pseudopotential construction is in no way different from the non-relativistic case.  However if a Dirac-Slater all-electron calculation is used as a basis for the the pseudopotential construction, there exists a problem of representing the  the up and down relativistic wave functions  and eigenvalues by a single wave function and eigenvalue arising from the non-relativistic Schr\"odinger equation.  Following the suggestion of  Kleinman,~\cite{Kleinman}  Bachelet and Schl\"uter~\cite{BS} proposed the construction of separate $V_{nl}^{\rm up}(r)$ and $V_{nl}^{\rm down}(r)$ for the up and down states. They then create an average pseudopotential weighted by the different $j$  degeneracies of the $l\pm \frac{1}{2}$ states 

\begin{eqnarray}
 V_{nl}^{\rm AVG}(r) = \frac {1}{2l+1} [l V_{nl}^{\rm down}(r) + (l+1) V_{nl}^{\rm up}(r)].
\end{eqnarray}

 The Bachelet-Schl\"uter (BS) averaged pseudopotential  contains all scalar parts of the relativistic pseudopotential.  This approach gives rise to an error of order $\alpha^{2}$, where $\alpha$ is the fine structure constant ($\alpha = \frac {1}{137.04}$). However the  eigenvalues of the BS averaged pseudopotential will not be equal to the weighted average of the up and down all-electron eigenvalues.  In addition, the valence charge density due to the solutions of the Schr\"odinger equation for the BS averaged pseudopotential will not be the same as the valence charge density of the all-electron atom.  Therefore, exact agreement between  the pseudopotential and the all-electron  orbital eigenvalues and charge density past $r_{c}$ is not preserved even for the  reference configuration.  Typically the error in the eigenvalues and total energies introduced by the potential averaging is small, about 1-4 mRy.  None of the modern pseudopotential methods can remove this error.   Thus one currently has a choice of either accepting the error due to  ignoring the spin-orbit coupling in the  all-electron calculations (Koelling-Harmon), or using the more correct Dirac-Slater equations and accepting the error due to pseudopotential  averaging.

We present below a method for constructing a pseudopotential whose eigenvalues agree exactly with the averaged relativistic all-electron eigenvalues

\begin{eqnarray}
 \epsilon_{nl}^{\rm AVG} = \frac {1}{2l+1} [l \epsilon_{nl}^{\rm down} + (l+1) \epsilon_{nl}^{\rm up}].
\end{eqnarray}

\noindent It is impossible to enforce charge density agreement with the relativistic all-electron calculation for all $r>r_{c}$, since each  wave function asymptotes to an exponential, and the  sum of two exponential functions cannot be equal to a single exponential function at more than two points.  Instead, we impose an aggregate norm conservation outside the core radius $r_{c}$.  We require the weighted average of the integrals  of the squares  of the spin-up and spin-down wave functions from $r_{c}$ to infinity  to be preserved by the new potentials and pseudo-wave functions for each valence orbital $\phi_{nl}$

\begin{eqnarray}
Q_{nl}&=&\int_{r_c}^{\infty} \left|\phi_{nl}(r)\right|^2 r^2dr \nonumber \\
      &=&\frac{1}{2l+1}\left(l\int_{r_c}^{\infty}\left|\psi^{\mathrm down}_{nl}(r)\right|^2 r^2 dr \right. + \nonumber \\
      & & \mbox{}(l+1)\left.\int_{r_c}^{\infty}\left|\psi^{\mathrm up}_{nl}
(r)\right|^2 r^2 dr\right).
\end{eqnarray}

 Unlike the standard averaging procedure where the spin-up and spin-down $pseudopotentials$ are averaged and the averaged pseudopotential is then used to obtain the non-relativistic  wave function and eigenvalues, in our procedure the relativistic all-electron  eigenvalues and the aggregate norm conservation criterion are used to obtain the non-relativistic wave functions and potentials at the all-electron level.  

The algorithm proceeds as follows. First we obtain the relativistic spin-up and spin-down all-electron wave functions and eigenvalues using the Dirac-Slater formalism.  The averaged all-electron eigenvalue and norm, $\epsilon_{nl}^{\rm AVG}$ and $Q_{nl}$, are then computed for each valence orbital.  The initial guess for the new non-relativistic $\phi_{nl}$ is set to be  the weighted average of the spin up and spin down Dirac-Slater wave functions $\psi_{nl}^{up}$ and $\psi_{nl}^{down}$, scaled to preserve $Q_{nl}$.  The valence charge density $\rho_{\rm valence}(r)$ corresponding  to  $\psi_{nl}^{up}$ and $\psi_{nl}^{down}$ is computed and then added to the charge density due to the core orbitals $\rho_{\rm core}(r)$ to give total charge density $\rho_{\rm total}(r)$.

The exchange-correlation and Hartree potentials and energy densities  are then computed  and the full potential $V(r)$ is obtained. 
For each orbital the potential $V(r)$ is used in an inward solve~\cite{Froese} for the new wave function $\phi_{nl}$ with the criterion $\epsilon_{nl}$ = $\epsilon_{nl}^{\rm AVG}$.  Each orbital is scaled so that the norm outside $r_{c}$ is  equal to $Q_{nl}$.  For $r < r_{c}$,   $\phi_{nl}$ is given a smooth nodeless form

\begin{eqnarray}
\phi_{nl}^{\rm AVG}(r) = r^{l+1} 
\left(\frac {\phi^{\rm AVG}_{i}(r_{c})}{r_{c}^{l+1}} + c_{nl} \left(1-\frac {r}{r_{c}}\right)^4 \right)
\end{eqnarray}

\noindent with the parameter $c_{nl}$ chosen such that $\phi_{nl}$ is normalized to unity. This is done for each valence orbital. The new valence charge density $\rho_{\rm valence}(r)$ is then computed from the $\phi_{nl}$, completing the cycle.  This cycle is repeated until the input and output $\phi_{nl}$ are equal.

\begin{table}[t]
\caption{Configuration testing for the Bachelet-Schl\"uter (BS) Au pseudopotential averaging and our all-electron averaging methods described in text.  Eigenvalues and eigenvalue errors are given for  all-electron (AE) weighted average eigenvalues,  BS averaged and our all-electron averaged pseudopotential eigenvalues  with and without the designed nonlocal (DNL) construction.  All energies are in Ry.}

\begin{tabular}{crrrrr}
 &\multicolumn{1}{c}{AE}&\multicolumn{1}{c}{BS}&\multicolumn{1}{c}{Present}&\multicolumn{1}{c}{BS} 
&\multicolumn{1}{c}{Present} \\
&\multicolumn{1}{c}{average}&\multicolumn{1}{c}{averaging}&\multicolumn{1}{c}{averaging}&\multicolumn{1}{c}{with DNL} 
&\multicolumn{1}{c}{with DNL} \\
&\multicolumn{1}{c}{}&\multicolumn{1}{c}{error}&\multicolumn{1}{c}{error}&\multicolumn{1}{c}{error} 
&\multicolumn{1}{c}{error} \\
\hline
& & & & & \\

$6s^1$&        	-0.4457&	 0.0011&	-0.0000&	 0.0011&	 0.0000\\
$6p^0$&        	-0.0696&	 0.0035&	 0.0000& 	 0.0035&	 0.0000\\
$5d^{10}$&       -0.5281&	 0.0035&	 0.0000&	 0.0035&	 0.0000\\
& & & & & \\

$6s^2$&           -0.5096&	 0.0013&	 0.0001&	 0.0018&	 0.0010\\
$6p^0$&           -0.0968&	 0.0050&	 0.0006&	 0.0050&	 0.0007\\
$5d^9$&           -0.6783&	-0.0027&	-0.0064&	 0.0000&	-0.0019\\
& & & & & \\

$6s^0$&           -0.9958&	-0.0008&	-0.0013&	 0.0009&	-0.0009\\
$6p^0$&           -0.5118&	 0.0038&	-0.0037&	 0.0036&	-0.0036\\
$5d^{10}$&          -1.1509&	 0.0037&	-0.0009&	 0.0041&	 0.0006\\
& & & & & \\

$6s^1$&           -1.0786&	 0.0010&	-0.0011&	 0.0016&	 0.0007\\
$6p^0$&           -0.5653&	 0.0059&	-0.0024&	 0.0052&	-0.0018\\
$5d^9$&           -1.3211&	-0.0031&	-0.0076&	 0.0005&	-0.0009\\
& & & & & \\

$6s^2$&           -1.1611&	 0.0009&	-0.0012&	 0.0025&	 0.0023\\
$6p^0$&           -0.6152&	 0.0081&	-0.0010&	 0.0064&	-0.0003\\
$5d^8$&           -1.5009&	-0.0099&	-0.0143&	-0.00183&	-0.0022\\
& & & & & \\

$6s^0$&           -1.6971&	 0.0022&	-0.0026&	 0.0007&	 0.0001\\
$6p^0$&           -1.0960&	 0.0052&	-0.0065&	 0.0015&	-0.0055\\
$5d^9$&           -2.0317&	-0.0041&	-0.0093&	 0.0001&	-0.0001\\
& & & & & \\

$6s^1$&           -1.7974&	-0.0018&	-0.0028&	 0.0012&	 0.0015\\
$6p^0$&           -1.1694&	 0.0074&	-0.0048&	 0.0010&	-0.0030\\
$5d^8$&           -2.2275&	-0.0111&	-0.0160&	-0.0021&	 0.0013\\	
& & & & & \\

$6s^2$&           -1.8947&	-0.0015&	-0.0026&	 0.0026&	 0.0033\\	
$6p^0$&           -1.2362&	 0.0100&	-0.0028&	-0.0010&	-0.0002\\
$5d^7$&           -2.4296&	-0.0154&	-0.0201&	 0.0002&	-0.0001\\
\end{tabular}
\end{table}


Pseudopotentials are then generated from the converged $\phi_{nl}$. The optimized  and designed nonlocal pseudopotential  methods are used in this paper to improve convergence  and  transferability.  However we emphasize that the present method of insuring agreement between relativistic all-electron and pseudopotential results can be used with any pseudopotential generating scheme, such as the ultrasoft construction, since the self-consistent solve  is done at the all-electron level.

We have applied our new averaging method to the gold atom. Gold has atomic number 79 and therefore must be treated relativistically.    We have used a neutral $6s^{1.0}$ $6p^{0.0}$ $5d^{10.0}$ reference configuration.  We use the Perdew-Zunger~\cite{PZ} parametrization of the Ceperley-Alder~\cite{CA} LDA for the exchange-correlation functional in all calculations.

As can be seen from the results in Table I, the new method  improves transferability in configurations close to reference, with the sum of absolute values of errors in orbital eigenvalues of neutral configurations reduced by 60$\%$.  Even more importantly, rigorously correct eigenvalues for the reference configuration enable the designed nonlocal method to improve transferability further.   Since the  DNL augmentation operator  does not affect the pseudo-wave functions and  eigenvalues of the reference configuration, the error given by the pseudopotential averaging method can $never$ be corrected with the designed nonlocal formalism.   Furthermore, the augmentation operator has less impact on configurations close to reference than on those further away; thus eigenvalue errors in the neutral configurations are much harder to eliminate for the pseudopotential  averaging method than for our new all-electron averaging method.  While the total transferability error for the new all-electron averaging based designed nonlocal potential is 40$\%$ less than that of  the BS averaging based designed nonlocal potential, the error in neutral configurations is 78$\%$ less.

We have presented in this paper a more accurate method for construction of relativistic pseudopotentials  based on Dirac-Slater all-electron atomic  calculations. Our procedure enforces the agreement of the pseudopotential and the all-electron spin averaged eigenvalues at the reference configuration, leading to a significant improvement in pseudopotential transferability. The method has negligible computational cost, is easy to implement, and applies to both norm-conserving and ultrasoft pseudopotential formalisms.

This work was supported by NSF grant DMR 97-02514 and Alfred P. Sloan Foundation  as well as the Laboratory for Research on the Structure of Matter and the Research Foundation at the University of Pennsylvania.  Computational support was provided by the San Diego Supercomputer Center and the  National Center for Supercomputing Applications.

\end{document}